\documentclass[11pt,twoside]{article}
\usepackage{asp2010}
\resetcounters
\aspvolume{??}
\aspcpryear{2010}
\aspvoltitle{UP: Have Observations Revealed a Variable Upper End of the Initial Mass Function?}
\aspvolauthor{M.\ Treyer, J.C.\ Lee, M.H.\ Seibert, T.\ Wyder, \&\ J.\ Neil, eds.}

\begin{document}

\title{Upper End IMF Variations Deduced from HI-Selected Galaxies}
\author{Gerhardt R. Meurer$^1,^2$
\affil{$^1$International Centre for Radio Astronomy Research, The University of Western Australia, M468, 35 Stirling Highway, Crawley WA, 6009 Australia\\
$^2$The Johns Hopkins University, Department of Physics and Astronomy, Baltimore MD, 21218 USA}}

\begin{abstract}
  Much of our understanding of modern astrophysics rest on the notion
  that the Initial Mass Function (IMF) is universal.  Our observations
  of a sample of HI-selected galaxies in the light of H$\alpha$ and the
  far-ultraviolet (FUV) challenge this result.  The flux ratio $F_{\rm
    H\alpha}/f_{\rm FUV}$ from these star formation tracers shows strong
  correlations with surface-brightness in H$\alpha$ and the $R$ band:
  Low Surface Brightness (LSB) galaxies have lower $F_{\rm
    H\alpha}/f_{\rm FUV}$ ratios compared to High Surface Brightness
  (HSB) galaxies as well as compared to expectations from equilibrium
  models of constant star formation rate (SFR) using commonly favored
  IMF parameters.  I argue against recent claims in the literature that
  attribute these results to errors in the dust corrections, the
  micro-history of star formation, sample issues or escaping ionizing
  photons.  Instead, the most plausible explanation for the correlations
  is the systematic variations of the upper mass limit and/or the slope
  of the IMF.  I present a plausible physical scenario for producing the
  IMF variations, and suggest future research directions.

\end{abstract}

\section{Introduction}\label{s:intro}

The assumption that the IMF is constant is commonly adopted in
astrophysics.  The basis for this assumption can be summarized as
follows.  Color-magnitude diagram (CMD) analyses of populous star
clusters, where the IMF is easiest to measure, show that the upper end
of the IMF has a slope close to the \citet{salpeter55} value $\gamma =
-2.35$, with cluster to cluster variations explainable by stochastic
variations in the number of stars in each cluster \citep{kroupa01}.
\citet[][hereafter LL03]{ll03} find that almost all star formation occurs in
star clusters.  Since clusters have a constant IMF and all stars form in
clusters, then the IMF must be constant. 

Recent studies \citep[including][]{hg08,lee+09} indicate problems with
this assumption. Our contribution to this field, \citet[][hereafter
M09]{meurer+09}, has been particularly forthright in its challenge to a
constant IMF.  Here I review results from that study, present a scenario
for the origin of the IMF variations and respond to critiques of our
paper.

\section{The SINGG and SUNGG surveys}\label{s:survs}

Our results are based on two surveys: the Survey of Ionization in
Neutral Gas Galaxies \citep[SINGG;][hereafter M06]{meurer+06} and it's
sister Survey of Ultraviolet emission in Neutral Gas Galaxies
\citep[SUNGG;][]{wong07}.  The parent sample of both of these is the HI
Parkes All Sky Survey \citep[HIPASS][]{meyer+04,koribalski+04}.  Nearly
500 galaxies were selected from HIPASS for SINGG so as to evenly
populate the HI mass function with preferentially the nearest galaxies
selected.  Over 300 of these were imaged in H$\alpha$ and the $R$ band.
Nearly 150 galaxies from SINGG were selected to be observed by GALEX in
the FUV and NUV bands for SUNGG.  104 galaxies with complete
optical and UV datasets were used in M09.

\begin{figure}
\plotone{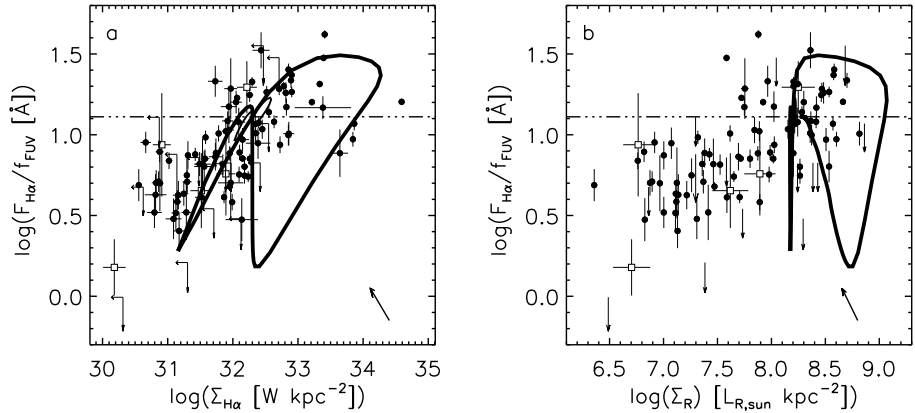}
\caption{Our main result: $F_{\rm H\alpha}/f_{\rm FUV}$ correlates with
  optical surface brightness, $\Sigma$. Panels (a), (b) plot the effective
  surface brightness in H$\alpha$ and $R$ band respectively.  
  Downward and leftward pointing arrows indicate cases with
  H$\alpha$ signal to noise ratio $< 2$.  The diagonal arrow in the
  lower right corner of each panel represents the average effect of dust
  absorption, which has been removed.  The broken horizontal line shows
  the $F_{\rm H\alpha}/f_{\rm FUV}$ value expected for a constant SFR
  population having a Salpeter IMF
  extending up to 100${\cal M}_\odot$. The curves represent models for a ``burst'' (SFR
  increase - loops to the right) or ``gasp'' (SFR decrease - loops to
  the left) SFH.  The SFH in each case has a constant base level with
  the burst or gasp having a Gaussian
  shape with fixed FWHM duration of 10 Myr.  Two
  cases for the ratio of maximum to minimum SFR for each SFH are shown
  2:1 and 100:1.  Additional
  models are shown in M09.
  \label{f:hafuvsb}}
\end{figure}

Our simple HI selection method means we do not suffer any of the common
optical biases.  This is illustrated in M06 where we show that all HI
targets in our first data release correspond to H$\alpha$ emitting
galaxies (in some cases multiple galaxies), and that all known
morphologies of star-forming galaxies are represented.  Similarly
\citet{hanish+06} demonstrate how our selection allows us to recover the
cosmic SFR density of the local universe.  These results
demonstrate that our sample is not biased against any particular type of
galaxy as long as it has HI.  Hence, we can rule out selection effects as
causing our results; this is an essential component of our claim that we
found true IMF variations (see \S\ref{s:sfh} below).

The ability to probe the IMF with two star-formation tracers was a major
design feature of our surveys.  H$\alpha$ emission traces the
distribution of the ionizing O stars, with initial masses down to ${\cal
  M}_\star \sim 20{\cal M}_\odot$, while the vacuum UV emission detected
by GALEX traces both O and B stars (initial masses $3{\cal M}_\odot
\la {\cal M}_\star \la 20{\cal M}_\odot$).  The total luminosity of a
star forming population is dominated by the vacuum UV and is fairly
insensitive to the parameters governing the upper end of the IMF.  The
fraction of ionizing light emitted, on the other hand, does depend
strongly on the IMF parameters.  Thus the ratio of H$\alpha$ to UV
emission is very sensitive to the IMF.

\section{The Results}\label{s:res}

Our key result is shown in Fig.~\ref{f:hafuvsb} - the ratio of H$\alpha$
line flux to FUV flux-density $F_{\rm H\alpha}/f_{\rm FUV}$ varies
strongly with optical surface brightness.  In these plots, each point
represents a single galaxy.  The $x$-axis gives the face-on effective
surface brightness $\Sigma$, while $F_{\rm H\alpha}/f_{\rm FUV}$,
plotted on the $y$-axis, corresponds to the ratio integrated over an
aperture large enough to encompass the entire galaxy.  By using integrated
quantities we avoid concerns about small-scale structure and evolution;
it doesn't matter that ionizing photons may escape HII regions, nor that
clusters dissolve, as long as the stars remain in the galaxy and the
ionization happens in the galaxy.  We find that $F_{\rm H\alpha}/f_{\rm
  FUV}$ also correlates with other quantities including morphology,
luminosity and dynamical mass.  The luminosity correlation was
independently found in the 11HUGS sample as discussed by \citep{lee+09}.
Since $\Sigma$ correlates with luminosity \citep{fs88,kauffmann+03b}, it
is not surprising that $F_{\rm H\alpha}/f_{\rm FUV}$ correlates with
both quantities. We find that the correlations with $\Sigma$ are
stronger and hence will concentrate on them (especially the $\Sigma_R$ correlation).
The LSB galaxies are most noteworthy in these
panels.  These are typically dwarf irregular (dI) galaxies.  Their low
$F_{\rm H\alpha}/f_{\rm FUV}$ ratios suggest a truncated IMF or steep
$\gamma$.  Next, I examine other factors that can affect $F_{\rm
  H\alpha}/f_{\rm FUV}$.  By showing that these can not account for our
results, we are left with IMF variations as the most likely explanation
for the observed correlations.

\subsection{Dust Attenuation}

\begin{figure}
\plottwo{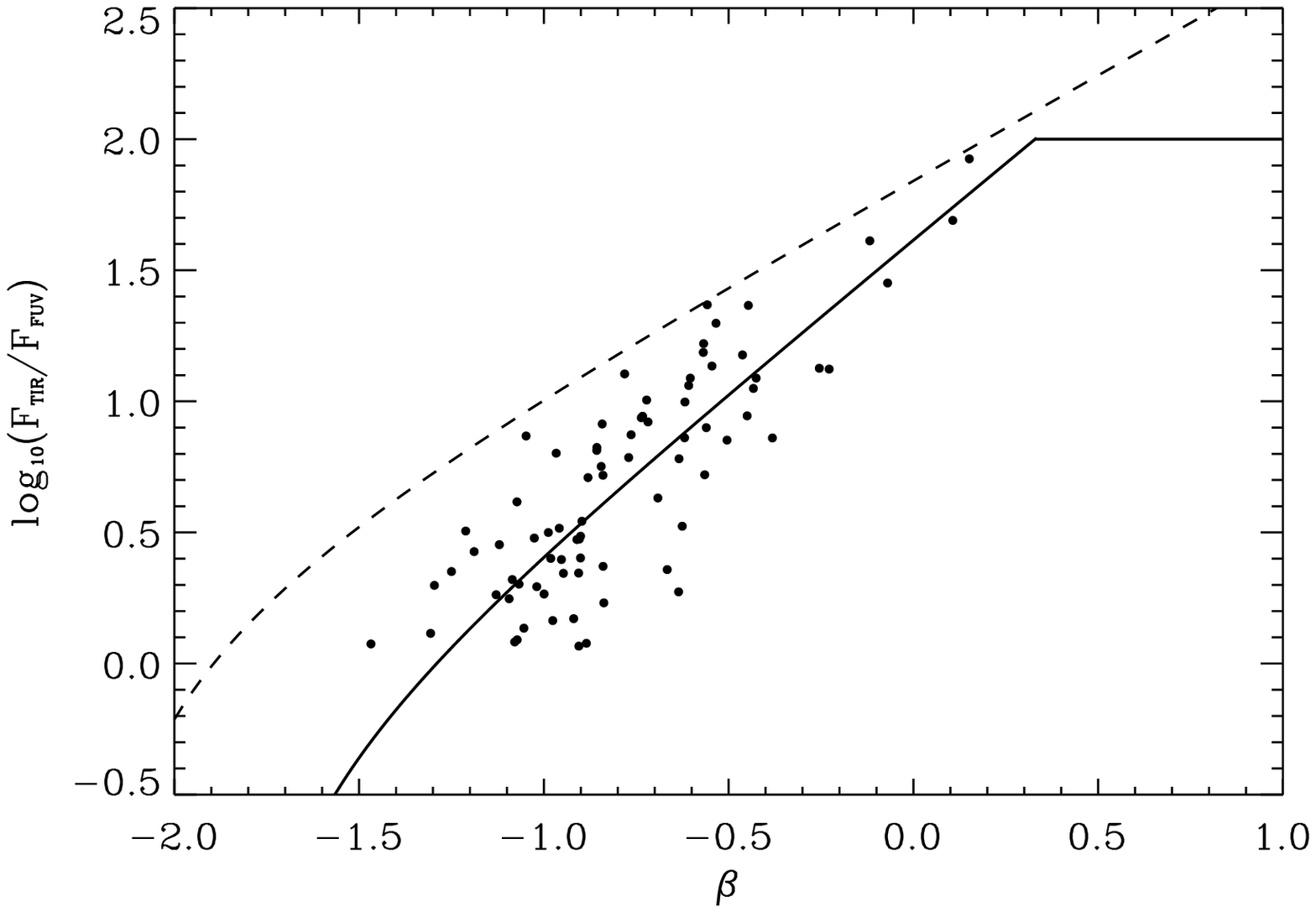}{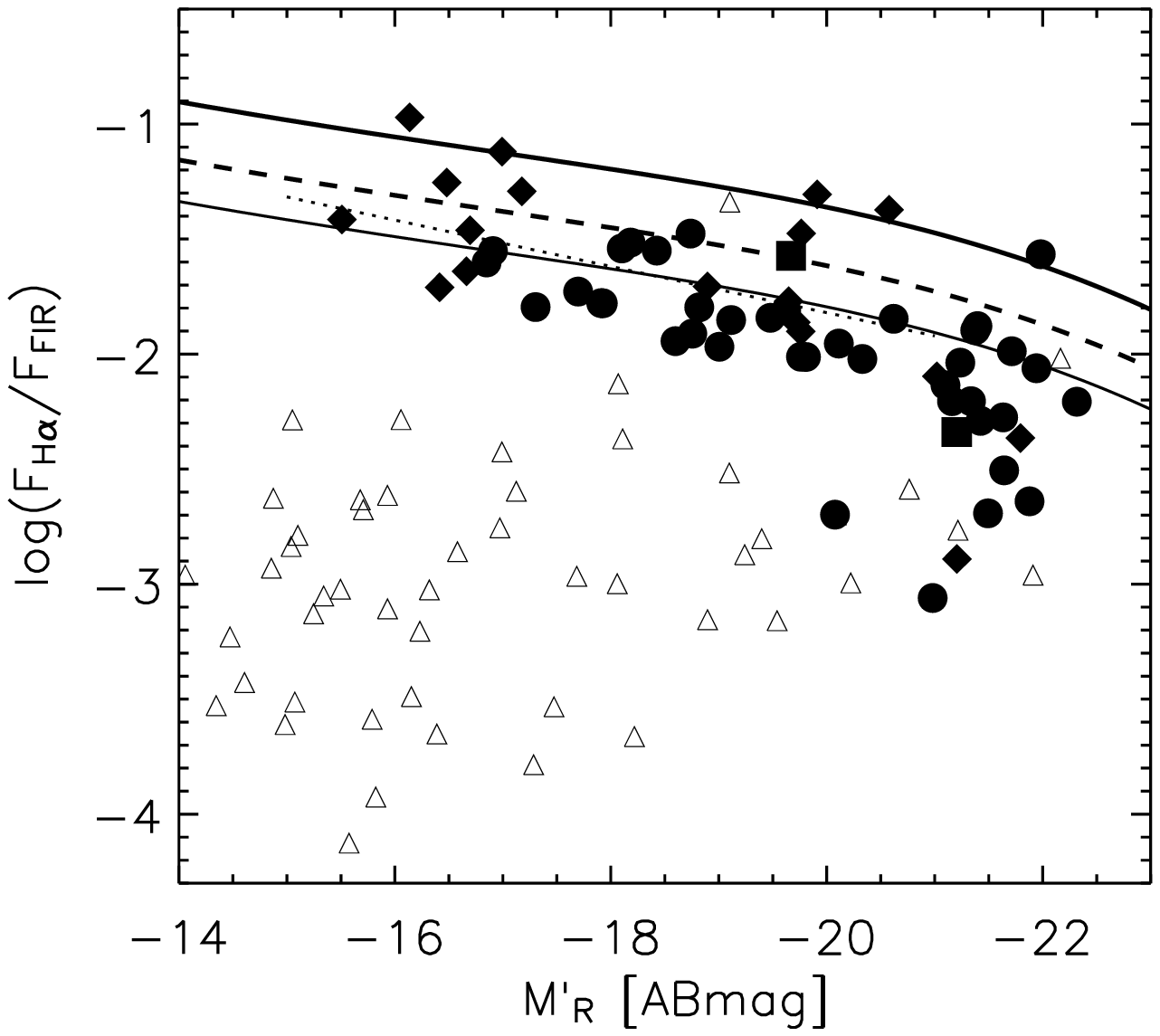}
\caption{Tests of our dust absorption and re-emission model.
  Panel (a) (left) shows the ratio of total far infrared emission to
  that in the FUV plotted against the ultraviolet spectral slope, $\beta$,
  (derived from the FUV-NUV color) for the SUNGG galaxies also detected
  in the FIR by IRAS \citep{wong07}.  Our
  calibration of the global dust reddening relation is shown as the
  solid line.  The calibration for starburst galaxies by \citet{mhc99}
  is shown as the dashed line.  Panel (b) shows the ratio of H$\alpha$
  and FIR fluxes plotted against $R$ band absolute magnitude.  Filled
  symbols mark IRAS detections, 
  while triangles indicate IRAS non-detections as $3 \sigma$ lower limits in $F_{\rm H\alpha}/f_{\rm FIR}$.  The curves
  represent the application of a simple dust reprocessing models on
  stellar population models.  The solid line is for a Salpeter IMF over a mass range of 1 - 100 ${\cal M}_\odot$;
  the dashed line is for the same $\gamma$ but over the mass range of 1 to 30
  ${\cal M}_\odot$.  The thin solid line and dotted line segment show fits to the
  data: the thick solid line shifted
  vertically and a simple linear fit respectively.  Further details can
  be found in M06.
  \label{f:dust}}
\end{figure}

A key concern with any UV or optical study 
is dust attenuation.  Our corrections for dust in the UV are based on
the IRX-$\beta$ relation for SUNGG galaxies shown in
Fig.~\ref{f:dust}a.  Our calibration falls below the relationship found
for starburst galaxies by \citep{mhc99}, and is similar to the
relationship found in other samples of local normal galaxies
\citep[e.g.][]{seibert+05,gildepaz+07}.  In the optical, we employ a
secondary correlation derived by \citet{hwbod04} from integrated Balmer
decrements measured in the Nearby Field Galaxy Survey \citep{jffc00}.
Figure~\ref{f:dust}b (from M06, reproduced by permission of the AAS) shows that our optical dust correction
can account for the relationship between the ratio of H$\alpha$ and FIR
fluxes and the $R$ luminosity for the SINGG galaxies detected by IRAS.
Fig.~\ref{f:dust} demonstrates that our dust corrections can account for
the attenuation that ends up in warm dust emission.  

 The key arguments against dust causing our correlations are that (a)
the typical correction ends up to be small
compared to the scale of the correlations (as shown by the reddening
vectors in Fig.~\ref{f:hafuvsb}),
and (b) the correlations between $\Sigma$ and $F_{\rm H\alpha}/f_{\rm
  FUV}$ exist even without any dust corrections.  This was shown in
Fig.~6 of M09, where the LSB end of the correlation is significantly
below the fiducial $F_{\rm H\alpha}/f_{\rm FUV}$ level expected for a
fully populated Salpeter/Kroupa IMF even before dust correction.  Using
a larger dust correction will move these galaxies further from the
fiducial ratio, not closer.

\citet[hereafter B09]{boselli+09} tested dust corrections similar in
form to ours.  This produced stronger correlations of $F_{\rm
  H\alpha}/f_{\rm FUV}$ with other quantities in their sample than when
they performed no corrections.  They argued that this is not physically
plausible and hence that such ``statistical'' corrections are unreliable
and should not be used.  While this is a legitimate concern, their
problem does not affect our study.  We find that the correlations after
dust corrections are somewhat worse than the uncorrected correlations,
as expected since the corrections are noisy and determined from
independent quantities.  Our derived average dust attenuation vector
points in nearly the same direction as the relationship derived by
\citet[][see Fig.\ 1 of M09]{calzetti01}, which is satisfying
considering how coarse the corrections are.  B09 go on to remove sources
that do not have FIR fluxes from their sample.  This results in a much
weaker correlation and very few galaxies with low $F_{\rm
  H\alpha}/f_{\rm FUV}$.  Dwarf galaxies, typically LSB, have very weak
FIR emission even in deep Spitzer observations \citep{lee+09,dale+09}.
By excluding galaxies without FIR emission, B09 are effectively throwing
out the most interesting part of their sample.

\subsection{Star Formation History (SFH)}\label{s:sfh}

Rapid changes in the SFR can lead to large changes
in $F_{\rm H\alpha}/f_{\rm FUV}$.  We used Starburst99 models
\citep{leitherer+99} to test whether SFR variations could cause the
observed correlations.  Our models consist of a population forming stars
at a constant rate with a Salpeter IMF over the mass range 0.1
to 100 ${\cal M}_\odot$ which then experiences a sudden increase (burst)
or decrease (gasp) in the SFR.  Further details of the model are given
in M09.  The results of the modeling are shown in
Fig.~\ref{f:hafuvsb}.  Panel (a) shows that strong changes in the SFR
can result in large simultaneous variations in both $F_{\rm
  H\alpha}/f_{\rm FUV}$ and $\Sigma_{\rm H\alpha}$.  However, panel (b)
shows that the
effects on $\Sigma_R$ are much smaller; neither the beginning of a gasp
nor the end of a burst can give the correlated decrease in both
$\Sigma_R$ and $F_{\rm H\alpha}/f_{\rm FUV}$ needed to explain low
$F_{\rm H\alpha}/f_{\rm FUV}$ galaxies.  Since HI selection does not
preclude LSB galaxies with high $F_{\rm H\alpha}/f_{\rm FUV}$ 
(nor HSB - low $F_{\rm H\alpha}/f_{\rm FUV}$ galaxies), we can not hide
galaxies from our selection during inopportune phases of the star
formation cycle.  If SFH causes the low $F_{\rm H\alpha}/f_{\rm FUV}$
cases, then all the LSB galaxies in nearby universe are simultaneously
going through a sharp decrease in the SFR.  Such synchronicity is
implausible and allows us to rule out SFH as the cause of the observed
correlations.  Similarly, \citet{hg08} find that the spread of H$\alpha$
EW versus optical colors in SDSS galaxies can only be explained by a
bursty SFH if there is an implausible synchronicity in the timing of the
bursts, and thus ruled out a variable SFR as causing their results.

Boselli (at this conference and in B09) argues that variations in the
``micro-history'' of star formation can explain the large range of
$F_{\rm H\alpha}/f_{\rm FUV}$.  However, his toy-model only considers
the ratio $F_{\rm H\alpha}/f_{\rm FUV}$ and not surface brightness.
This does not address the correlations which are at the heart of the
problem. Instead he suggests the experiment of considering a dwarf
galaxy with and without the 30 Dor region.  The actual model used in B09
consists of regularly spaced bursts with no star formation between
bursts.  The inter-burst period they require to produce the $F_{\rm
  H\alpha}/f_{\rm FUV}$ distribution would also produce many H$\alpha$
non-detections which is inconsistent with SINGG results
(M06).  The dwarf galaxies usually associated with bursts
are Blue Compact Dwarfs (BCD).  \citet{mhc99} show that the bursts in
BCDs amount to at best a factor of a few change in the SFR.  Thus an
appropriate model for the SFH in dwarf galaxies is the factor of two
burst/gasp model, shown as the thin lines in Fig.~\ref{f:hafuvsb}, which
are barely discernible in these plots because the excursions in $F_{\rm
  H\alpha}/f_{\rm FUV}$ versus $\Sigma$ are relatively small.
Hypothetical extreme bursts are required to explain the $F_{\rm
  H\alpha}/f_{\rm FUV}$ excursions, and even then, the correlation with
$\Sigma_R$ is not explained as noted above.

\subsection{Escaping Ionizing Radiation}

Using H$\alpha$ as a star forming indicator implicitly assumes that all
ionizing photons are absorbed by the ISM.  The observed $F_{\rm H\alpha}/f_{\rm FUV}$
correlations could occur if the escape fraction of ionizing photons,
$f_{\rm esc}$, inversely correlates with $\Sigma$.  This scenario would
require $f_{\rm esc} > 0.5$ to explain the galaxies in the lowest
quartile of $F_{\rm H\alpha}/f_{\rm FUV}$.  While this is a large
effect, no attempt has been made to
observe emission with a rest $\lambda_0 < 912$\AA\ in LSB galaxies, so
it is not yet possible to formally rule out this scenario.  Escaping
ionizing photons are the preferred explanation of the low $F_{\rm
  H\alpha}/f_{\rm FUV}$ ratios given by \citet{hel10}.  They also imply
that some H$\alpha$ is missed from the observations because it is below
the $\Sigma_{\rm H\alpha}$ detection limit.

The detection limit concern is easily dismissed.   Our measurements and
upper limits are state of the art and accurate to within the error bars shown, even when
H$\alpha$ is not detected in some individual pixels, because those
pixels are within the large elliptical apertures employed in our
analysis (M06).  A large $f_{\rm esc}$ also seems unlikely for two reasons.  First, the problem
cases are the LSB galaxies, and these typically have a higher ISM
content (larger ${\cal M}_{HI} / L_B$) and puffier disks than the larger
spiral galaxies.  Hence, they should be able to retain their ionizing
photons better.  Second, CMDs of the nearest galaxies indicate that LSB
galaxies typically are deficient in O stars \citep[e.g.][]{yswd07}
compared to HSB irregulars and BCDs \citep[e.g.][]{annibali+08}.  Since
such studies usually have been on a galaxy per paper basis, then the 
assumption of a fully sampled Salpeter IMF can fool any team of
astronomers to find that their particular galaxy has a gasping SFH. A
better approach would be to obtain uniform high quality HST data of many
nearby galaxies and then compare CMDs as a function of $\Sigma$.  If the
sample is large enough, galaxy to galaxy variations will average out and
an effective IMF can be deduced.  The ANGST team have obtained such a
dataset \citep{dalcanton+09} and are well poised to do such a study.

\section{What Causes Upper End IMF Variations?}\label{s:disc}

\subsection{Cluster Versus Field Star Formation}

As noted in \S\ref{s:intro}, much of the support for a constant IMF
rests on the notion that all star formation occurs in star clusters.
However, H$\alpha$ and UV images of galaxies show the majority of the
light is diffuse rather than being in discernible clusters or HII regions.
With SINGG we find that on average $\sim 40\%$ of the H$\alpha$ emission
is in HII regions while the rest is diffuse \citep{Oey+07}.  In UV and
$U$ band light, the fraction of light that is in compact clumps
(presumably clusters) is typically $\sim 20$\%\ for starbursts
\citep{meurer+95} and only a few percent for normal galaxies
\citep{lr00}; the vast majority is diffuse.  The dominant diffuse
population has low $F_{\rm H\alpha}/f_{\rm FUV}$ \citep{hwb01} and a UV
spectrum dominated by B stars, whereas the young ``super star clusters''
typically have O star rich spectra \citep{tclh01}.  The disparity
between cluster and field is typically ascribed to infant mortality
\citep[e.g.][]{tclh01,pellerin+10}.  The main problem with this scenario
is that naiveley one would expect infant mortality to be stronger in the
dense environment of starbursts than in normal and LSB galaxies, whereas
the results of \citet{meurer+95} and \citet{lr00} indicate that clusters
are {\em more\/} prevalent and thus last longer in starbursts.

The interpretation from LL03 that all star formation occurs in clusters
is based on an expansive redefinition of the term ``cluster'' to include
objects that are unbound at birth (when they shed their natal gas).
Better terms for these unbound objects are the older terms ``group'' and
``association'', which now seem to be largely neglected by the
extragalactic astronomical community.  However, it is important to
retain such a distinction, because it is crucial in the physics of high
mass star formation.  Simulations indicate that high
mass stars can form readily in dense {\em bound\/} clusters where
proto-stars can quickly build-up to high mass by competitive accretion
\citep{bbv03,bvb04}.  Alternatively, \citet{km08} show that high mass
stars can form efficiently in a top-down fashion when the ISM density is
very high, because then fragmentation is inhibited. In either scenario,
the physics is different for star formation in dense clumps compared to
low density or unbound objects. Recent large area infrared surveys
reveal a significant ``distributed mode'' of star formation comprising
$\sim30$\%\ of star formation in the solar neighborhood
\citep{allen+07,mln09}.  This star formation, combined with the large
fraction of unbound objects that LL03 refer to as clusters, is likely to
make the majority of the diffuse UV emission in galaxies.

\subsection{Our Scenario}

We posit that the $F_{\rm H\alpha}/f_{\rm FUV}$ ratio variations arise
from the difference between star formation in bound clusters and the field.  Bound
clusters need a dense cold molecular ISM to form.  The fraction of the
total ISM that goes into this phase is set by the hydrostatic pressure
\citep{mo77,br06}, which also determines how tightly bound clusters are
when they form \citep{ee97,elmegreen08}.  Stars dominate the disk plane
potential, thus largely setting the pressure.  This results in the
correlation with $\Sigma_R$ which also measures the disk plane mass
density.  A high mass density disk results in high pressure ISM, with a
larger fraction of its mass in cold dense molecular clouds.  These form
dense bound clusters that are rich in massive O stars, and hence a high
$F_{\rm H\alpha}/f_{\rm FUV}$.  Conversely, a low mass density disk will
have lower $\Sigma$, lower pressure, and less formation of
bound clusters relative to field stars and unbound
objects.  This results in a low $F_{\rm H\alpha}/f_{\rm FUV}$.

\subsection{Implications}

A variable IMF has numerous major implications.  Star formation surveys
that adopt a constant IMF are likely to get a biased result depending on
the tracer they use and the pressure of the ISM.  UV based SFRs should
be more accurate than those derived from H$\alpha$ \citep{lee+09}.
Similarly, a variable IMF throws in to doubt the basis for measuring the
SFH from CMDs using the tip of the main sequence as a clock.  Finally a
variable IMF provides an alternative explanation for the mass
metallicity relationship \citep{tremonti+04}.  Instead of requiring a
galactic wind to remove excess metals from dwarf galaxies, a truncated
IMF means that they may not have made them in the first place.

\section{Conclusions and Future Work}\label{s:conc}

The correlation between global $F_{\rm H\alpha}/f_{\rm FUV}$ and the
surface brightness of galaxies provides strong evidence that the IMF is
varying, and suggests that the underlying cause is the hydrostatic
pressure of the ISM which regulates the phase balance of the ISM and
consequently the efficiency of the highest mass stars to form.  As
discussed throughout this meeting (and noted above) a variable IMF has major implications for
many branches of astrophysics

While I believe that the preponderance of evidence clearly demonstrates
the variable nature of the IMF, the astronomical community, as a whole,
is not yet convinced. One common thread in studies showing IMF
variations is the use of H$\alpha$ as a star forming indicator.  It
would be good to confirm these results using other methods, such as CMD
studies. Deep CMD studies of large samples of nearby galaxies (e.g.\
ANGST) are also critical for determining how far down in mass the the
IMF variability goes. Such studies would also provide useful tests of
the IGIMF hypothesis \citep{wk05,pwk09} which makes predictions on the
form of the galaxy wide IMF.  Cases of high $F_{\rm H\alpha}/f_{\rm FUV}$, well
above Salpeter expectations, as seen in our work and that of
Gunawardhana (this meeting), are also interesting.  The highest values
may be pushing beyond what models can achieve by just adjusting the
upper mass limit or $\gamma$, and may instead require truncating the
lower end of the IMF at values above 3 ${\cal M}_\odot$.  Such a
short-wick IMF is similar to what is required at high redshift to
explain differences between the cosmic SFR density and the mass assembly
history \citep{dave08,wth08,whtt08}.  Hence high $F_{\rm H\alpha}/f_{\rm
  FUV}$ galaxies may be good analogs to high redshift galaxies.

\acknowledgements I thank the SINGG and SUNGG team for their work on
this project, especially Dan Hanish, Ivy Wong, Zheng Zheng, and Ji Hoon
Kim for their analysis work.  Support for this was obtained in part
through NASA Galex Guest Investigator grant GALEXGI04-0105-0009 and NASA
LTSA grant NAG5-13083.

\bibliography{meurer_g}

\end{document}